\documentstyle[12pt]{article}

\begin{document}
\pagestyle{empty}
\title{ \bf Quantum Mach effect by Sagnac Phase Shift on Cooper pairs in
rf-SQUID }
\author{D. Fargion$^{1,2}$, L. Chiatti$^3$, A. Aiello$^1$ \\
$^1$ Physics Department of "La Sapienza" University and $^2$INFN,
Rome \\ $^3$Medical Physics Laboratory ASL VT, Viterbo (Italy)}
\date {October 5, 1999}
\maketitle
 \thispagestyle{empty}
\baselineskip=30pt
\begin{abstract}
The inertial drag on Cooper pairs in a rotating rf-SQUID ring
leads to an asymmetry in the Hamiltonian potential able to modify
the tunnelling dynamics. This effect is a Sagnac signature of
Cooper pairs at a quantum level and it may probe the Mach effect
influence on the wavefunction collapse.
\end{abstract}
\newpage
\pagestyle{plain} \setcounter{page}{1}
\section{Introduction}
The Mach principle (that is, the overlapping of the ``inertial
absolute space" with the ``fixed star" or cosmic BBR inertial
frame) constitutes one of the most relevant open question in
general physics. It relates fundamental questions from
microphysics (the origin of the elementary particle inertial mass)
to general relativity (the Lense Thirring drag) as well as to
basic cosmology. At classical level the Mach principle links
different puzzling historical questions \cite{barbour}: from the
oldest Newton bucket and the Foucault's pendulum to recent
Sagnac's experiments (gravitomagnetism \cite{wein}) due to the
Lense and Thirring drag on gyroscopes \cite{ciu}. The Mach effect
is partially implied by General Relativity \cite{wein} but it
calls for a more stringent test at quantum level. A quantum system
probing the absolute rotation has been considered long time
before: it is a superfluid helium gyroscope (SHEG)
\cite{cer,pack}, which has recently been brought very close to
reality by Packard's group \cite{aven}. The effectively measured
physical quantity in a SHEG experiment is the rotation induced
helium flow, that is the value of a rotation dependent observable.
Our proposal is essentially different in principle because it aims
to measure the rotation dependent {\em  probability } of a given
result when the collapse of the system wavefunction occurs in a
Macroscopic Quantum Coherence (MQC) experiment. In other words,
while the SHEG explores the rotation influence at the level of
physical quantities, we investigate the perturbation on the
wavefunction collapse mechanism.
\section{The MQC experiment}
The  MQC experiment has extensively been discussed elsewhere
\cite{leg1,leg2} and here we limit ourselves to recall its
essential features. The rf-SQUID consists of a superconducting
ring having self-inductance $L$ in which a Josephson junction of
critical current $I_c$ and capacitance $C$  is inserted
\cite{barone}. The generalised four-momentum of Cooper pairs
circulating in the SQUID is expressed as:
\\
\begin{equation}\label{1}
 \Pi_\mu = p_\mu - 2 \frac{e}{c} A_\mu, \qquad \mu = 0,1,2,3 .
\end{equation}
\\
The integral of the generalised momentum along the rf-SQUID ring
must satisfy the Bohr quantization rule:
\\
\begin{equation}\label{2}
 \oint \Pi_i d x^i = n h + \alpha_n  \hbar, \qquad i = 1,2,3;
 \quad n = 0,1,2, \ldots,
\end{equation}
\\
where  $\alpha_n$  is  the phase shift difference at the edge of
the Josephson junction. The above equation for the Cooper electron
pairs (of mass and energy  $m_{\mathrm{{cp}}} \approx 2 m_e$,
$\varepsilon_{\mathrm{{cp}}} \approx 2 \varepsilon_e$), whose
kinematics momentum is $2 m_e \gamma u_i$  (where $\gamma$ is the
Lorentz factor related to the tangential velocity of the ring and
in non relativistic regime $\gamma \approx 1$), becomes ($\Phi_0
=  h / 2e$, $\Phi =$ magnetic flux threading the ring):
\\
\begin{equation}\label{3}
- \frac{2 m_e \gamma}{h} \oint u_i d x^i - \frac{\Phi}{\Phi_0} = n
+ \frac{\alpha_n}{2 \pi}, \qquad n = 0,1,2, \ldots,
\end{equation}
\\
where the minus sign in front of the Cooper pairs velocity $u_i$
is due to the fact that ``positive" currents (fixing the
integration versus) flows in opposite versus with respect to the
negative moving charges.  Assuming zero vorticity Eq. (\ref{3})
becomes:
\\
\begin{equation}\label{4}
 \alpha_n = - 2 \pi n - 2 \pi \frac{\Phi}{\Phi_0},
  \qquad n = 0,1,2, \ldots .
\end{equation}
\\
We remind that our present simple derivation agrees, in
non-relativistic regime, with the rigorous
 General Relativity treatment of rotating superconducting ring.
 Indeed following ref. \cite{jee1} (formula $\mathbf{(3.3)}$), one finds  a
generalized ``London moment'' as the following one
\\
\begin{equation}\label{25}
\frac{ 2 e}{\hbar c} \int_\Sigma {\mathbf{B}} \cdot  {\mathbf{d
s}} + \frac{ 2 }{\hbar c^2} \zeta \oint {\mathbf{u}} \cdot
{\mathbf{d r}} = 2 \pi n,
\end{equation}
\\
where $\zeta \approx m c^2$ in  non-relativistic regime. The
explicit integral on a superconducting circular ring may be
written as:
\\
\begin{equation}\label{26}
e B_0 \Sigma + 2 e V \frac{\Omega \Sigma}{c} + 2 \zeta
\frac{\Omega \Sigma}{c} = n \frac{h c}{2},
\end{equation}
\\
where the second term is proportional to the electric field
induced from the magnetic one made by Lorentz boost, and it can
always be  set, for the non-relativistic regime under
consideration,to a negligible (or a vanishing) value with respect
to the electron rest mass ($e V \ll \zeta $). The third term,
related to the particle (Cooper pair) angular momentum, in a
non-relativistic regime reduces exactly (in CGS units), to our Eq.
(\ref{3}) where it is present the additional well know phase shift
\cite{papini} due to the presence of the  Josephson junction.
\\
The interaction energy due to the magnetic coupling of the
junction is \cite{barone}:
\\
\begin{equation}\label{5}
  -\frac{\Phi_0 I_c}{2 \pi} \cos (\alpha_n) =
  -\frac{\Phi_0 I_c}{2 \pi} \cos \left( 2 \pi \frac{\Phi}{\Phi_0} \right).
\end{equation}
\\
The junction capacity $C$ induces the accumulation of  the
electrostatic energy $C V^2 / 2$, where $V = d \Phi /dt $ is the
voltage across the junction. Introducing the ``momentum" $ P_\Phi
= C V $ conjugated to $\Phi$,  this energy can be written as  $
P_\Phi^2 / 2C $. Adding this expression to Eq. (\ref{5}) and to
the magnetic energy of the superconducting ring, leads to the
following Hamiltonian for a non rotating rf-SQUID :
\\
\begin{equation}\label{6}
\hat{H} = \frac{P_\Phi^2}{2 C} + \frac{(\Phi - \Phi_0/2)^2}{2 L} -
\frac{\Phi_0 I_c}{2 \pi} \cos (\alpha_n).
\end{equation}
\\
Through the usual replacement $ P_\Phi \rightarrow - i \hbar
\partial_\Phi $, the above equation becomes, for $n = 0$
\\
\begin{equation}\label{7}
\hat{H} = -\frac{\hbar^2}{2 C} \partial_\Phi^2 +
  \frac{ ( \Phi - \Phi_0/2 )^2 }{2 L}
  - \frac{\Phi_0 I_c}{2 \pi} \cos \left( 2 \pi
\frac{\Phi}{\Phi_0} \right),
\end{equation}
\\
where an external flux   $\Phi_{\mathrm{ext}} = \Phi_0 / 2$ has
been assumed. The potential is clearly symmetric around $ \Phi_0 /
2 $ and, for suitable choice of $L$ and $I_c$, it presents two
minima at values $\Phi_0 /2 \pm \Phi_1 $, corresponding to
opposite  versus of the current $ I = (\Phi - \Phi_0/2) / 2L $
circulating in the ring. The first two energy eigenvalues, say
$\varepsilon_S$ and $\varepsilon_A $, correspond  to Symmetric and
Antisymmetric solutions $\psi_S$ and $\psi_A$, respectively. The
linear combinations $\psi_L = \psi_S + \lambda \psi_A$,$\psi_R =
\psi_S - \lambda \psi_A$ represent two state centered around Left
and Right minimum respectively ; the sign of $\lambda $ determines
if the symmetric or the antisymmetric linear combination is left
centered ($\lambda =1$ in the first case ,$\lambda = -1$
otherwise). If we set $\psi = \psi_L$ at $t = 0$, the probability
difference $P(t)$ to observe $\Phi < \Phi_0 /2$ ($P_L$) with
respect to the probability to observe $\Phi > \Phi_0 /2$ ($P_R$),
by an ideal measure at any time $t$, is the usual function
\\
\begin{equation}\label{8}
  P(t) \equiv P_L(t) - P_R(t) = \cos (2 \pi \nu_{\mathrm{tu}}t),
\end{equation}
\\
where $\nu_{\mathrm{tu}}$ is the tunnelling frequency between the
two well for the ``rest" Hamiltonian:
\\
\begin{equation}\label{9}
\nu_{\mathrm{tu}} = \frac{\varepsilon_A - \varepsilon_S}{2 h}.
\end{equation}
\\
Any measurement resets the probability expectation value; if at
time $ t$ results $\Phi < \Phi_0/2$ than, for $t' > t$, Eq.
(\ref{8}) holds replacing $t$ with $t' - t$. In other words the
measurement induces a ``wave-packet collapse". We remind that this
discussion lies upon the assumption that the circuitation
appearing in Eq. (\ref{3}) is null, and $n = 0$.
\section{The Sagnac effect in rotating rf-SQUID}
The simplest classical Sagnac effect for a {\em massless
particle}, as a photon, in an oriented ring system of radius $R$
and surface $ |{\mathbf{\Sigma}}| = \pi R^2 $, rotating at angular
velocity ${\mathbf{\Omega}}$ (for instance in a ring laser or in a
fiberoptic ring), may be naively evaluated: the propagation time
$\tau$ along a circular path of radius $R$ is $\tau = 2 \pi R /
v$,  where  $v$ in general is the phase velocity of the photon in
the medium; it differs for co-rotating and counter-rotating
photons by a time deviation interval (for the ideal in axis
rotating ring)
\\
\begin{equation}\label{10}
\Delta t = \frac{(2 \pi + \Omega \tau)R}{v} - \frac{(2 \pi -
\Omega \tau)R}{v} = \frac{4 \pi R^2 \Omega}{v^2},
\end{equation}
\\
where $\Omega \equiv |{\mathbf{\Omega}}|$. In the general cases to
discuss we set $v \approx c$. The corresponding phase shift
$\delta \alpha$ between two counter-propagating modes (of a
monochromatic source) at a given frequency $\nu$ whose energy is
$\varepsilon_\gamma = h \nu$, in general becomes:
\\
\begin{equation}\label{11}
\delta \alpha = \frac{4 \varepsilon_\gamma}{\hbar c^2}
{\mathbf{\Omega}} \cdot {\mathbf{\Sigma}},
\end{equation}
\\
and it is therefore as usual proportional to the area and angular
velocity. For a counter-rotating and co-rotating wave packet of
any {\em massive} particle whose energy is $\varepsilon_p = m^*
c^2 \gamma$  (in non relativistic regime $\gamma \approx 1$), the
above Sagnac phase shift may be naturally extended \cite{sak}, by
substitution $\varepsilon_\gamma \rightarrow \varepsilon_p$ in
equation (\ref{11}), as follows:
\\
\begin{equation}\label{12}
\delta \alpha = \frac{4 \varepsilon_p}{\hbar c^2}
{\mathbf{\Omega}} \cdot {\mathbf{\Sigma}} = \frac{4 m^*
\gamma}{\hbar} {\mathbf{\Omega}} \cdot {\mathbf{\Sigma}}.
\end{equation}
\\
The classical detection of such a phase shift allows the observer
to be informed of its own rotation (for instance of the
terrestrial angular velocity $\Omega_\oplus$) with respect to the
``absolute inertial frame". As Mach noted this frame is coincident
with the wider cosmological one defined ``fixed stars" or, better,
in a cosmological language (because of the linkage between matter
and radiation at recombination), by the BBR inertial frame. This
peculiar coincidence, already verified at  $10^{-6}\Omega_\oplus$
level, call for a gravitational and/or cosmic root of the inertia.
In order to measure the Sagnac phase shift effect in rf-SQUID
system let us calculate the circuitation appearing in equation
(\ref{3}) when the rf-SQUID rotates around its central axis with
angular velocity ${{\Omega}}$. The average velocity of the Cooper
pairs in the rotating system is ${{\Omega}} R$ so that the
circuitation (or vorticity) is just the Sagnac phase shift $\delta
\varphi$
\\
\begin{equation}\label{13}
\delta \varphi = -\frac{2 m_e \gamma}{h} \oint u_i d x^i =
-2\frac{2 m_e\gamma}{h} {\mathbf{\Omega}} \cdot {\mathbf{\Sigma}},
\end{equation}
\\
and the equation (\ref{4}), for $n = 0$, becomes:
\\
\begin{equation}\label{14}
\alpha_0 = \delta \varphi - 2 \pi \frac{\Phi}{\Phi_0}=  -2\frac{2
m_e \gamma}{h} {\mathbf{\Omega}} \cdot {\mathbf{\Sigma}} - 2 \pi
\frac{\Phi}{\Phi_0}.
\end{equation}
\\
If the circuitation versus is inverted (counter-rotating current)
the phase shift in Eq. (\ref{14}) change sign. The difference
between the phase shift corresponding to co-rotating and
counter-rotating currents gives the Sagnac phase shift of Eq.
(\ref{12}) (where $m^* = 2 m_e$ is the mass of Cooper pair); Eq.
(\ref{14}) can be also derived from general relativity in a more
rigorous way \cite{jee1,jee2}. In presence of a rotation the Eq.
(\ref{5}) becomes, because of Eq. (\ref{14})
\\
\begin{equation}\label{15}
- \frac{\Phi_0 I_c}{2 \pi} \cos (\alpha_n) = - \frac{\Phi_0 I_c}{2
\pi} \left| \cos \left( 2 \pi \frac{\Phi}{\Phi_0} \right) \cos
(\delta \varphi)  + \sin \left( 2 \pi \frac{\Phi}{\Phi_0} \right)
\sin (\delta \varphi) \right|.
\end{equation}
\\
Therefore the interaction term of the Hamiltonian (\ref{7})
becomes asymmetric with respect to the non rotating case. The last
asymmetric term in Eq. (\ref{15}) derives from the influence of
the Josephson junction motion on the Cooper pairs. The
perturbation to the Hamiltonian potential deforms the symmetric
``two well" into an asymmetric ``two well". The deepest minimum
will collect most of the tunnelling event states, leading to a
more probable configuration where the (negative) Cooper pairs are
more often counter-rotating the same rotating system. The
probability expressed in (\ref{8}) now becomes \cite{grab}
\\
\begin{equation}\label{16}
P(t) = \frac{1}{ 1 + \left(
\displaystyle{\frac{\nu_{\mathrm{tu}}}{\nu_\Omega}} \right)^2} +
\frac{\left( \displaystyle{\frac{\nu_{\mathrm{tu}}}{\nu_\Omega}}
\right)^2}{ 1 + \left(
\displaystyle{\frac{\nu_{\mathrm{tu}}}{\nu_\Omega}} \right)^2}
\cos \left( 2 \pi \sqrt{\nu_{\mathrm{tu}}^2 + \nu_\Omega^2} t
\right),
\end{equation}
\\
where
\\
\begin{equation}\label{17}
\nu_\Omega \equiv \frac{|\delta H_\Omega|}{h},
\end{equation}
\\
and $\delta H_\Omega$ is the energy gap between the two potential
minima. For $\delta \varphi \ll 1$, as it is in the most realistic
cases, the following approximation holds
\\
\begin{equation}\label{18}
\delta H_\Omega \approx - 2 \frac{\Phi_0 I_c}{h} \sin \left( 2 \pi
\frac{\Phi_1}{\Phi_0}  \right) \delta \varphi \approx - 2
\frac{\Phi_0 I_c}{h} \sin \left( 2 \pi \frac{\Phi_1}{\Phi_0}
\right) \left( \frac{2 \varepsilon_{\mathrm{cp}}}{\hbar c^2}
\right) {\mathbf{\Omega}} \cdot {\mathbf{\Sigma}} .
\end{equation}
\\
In a qualitative way we may imagine the Josephson junction playing
the role of a potential well for the two versus currents. While at
rest the transmission in both directions is symmetric, once the
rf-SQUID is rotating the nominal kinetic energy of the Cooper
pairs (in the rest frame of the junction) appears slightly
asymmetric: the higher energy of the current co-rotating respect
with to angular velocity (where the negative Cooper charges are
hitting the junction at higher velocities) leads to an easier
tunnelling probability through the Josephson junction. For the
same reason the counter-rotating current (where the Cooper pairs
are hitting at lower kinetic energy) has a higher energy gap and a
consequent lower transmission probability. Therefore one discovers
an ``anti-Lenz" law whose validity is based only on the negative
charge nature of Cooper pairs. For an ideal positive boson charge
the opposite will be true. It should be noticed that, for
${\mathbf{\Omega}} \cdot {\mathbf{\Sigma}} > 0$, because on the
right of Eq. (\ref{18}) $\Phi \geq \Phi_0 /2$, the final sign of
$\delta H _\Omega$ on that side is negative and the two well
potential will be unbalanced as follows: ``up" on the left and
``down" on the right of $\Phi_0 /2$. The physical meaning is that
the most probable state will be at ``positive" flux ($\Phi \geq
\Phi_0 /2$), i.e., the ``positive" ${\mathbf{\Omega}}$  rotation
will induce a flux and an average magnetic field along its vector
sign (as mentioned in the introduction) leading to a preferential
flux $\Phi \geq \Phi_0 /2$. We remark once again this anti-Lenz
result is due only to the negative charge nature of the Cooper
pairs. In the following we want to quantify the perturbation
effect in a realistic measure set up.
\section{The experimental set up}
In the non rotating rf-SQUID the position of minima is calculated
solving the equation $\partial U / \partial \Phi_0 =0$, $U$ being
the potential appearing in Eq. (\ref{7}). In the parabolic
approximation of the well bottom the ground state level energy is
given by $\varepsilon_0 = h \nu_{{LC}}/2$, where
\\
\begin{equation}\label{19}
\begin{array}{cl}
\nu_{{LC}} & = \displaystyle{\frac{1}{2 \pi}}
\sqrt{\displaystyle{\frac{1}{C} \frac{\partial^2 U}{\partial
\Phi^2} \bigg|_{\Phi = \Phi_0 /2 \pm \Phi_1 } }} \\\\ & =
\displaystyle{\frac{1}{2 \pi}} \sqrt{\displaystyle{\frac{1}{LC}}}
\sqrt{\displaystyle{ 1 + \frac{2 \pi L I_c}{\Phi_0} \cos \left( 2
\pi \frac{\Phi}{\Phi_0} - \pi \right)
 }}.
\end{array}
\end{equation}
\\
The tunnelling frequency is calculated as
\\
\begin{equation}\label{20}
  \nu_{\mathrm{tu}} \cong \nu_{LC} e^{- S/ \hbar},
\end{equation}
\\
where
\\
\begin{equation}\label{21}
\frac{S}{ \hbar} = \int_{\Phi_L (\varepsilon_0)}^{\Phi_R
(\varepsilon_0)} \sqrt{2 C \left[ U (\Phi) - \varepsilon_0
\right]} d \Phi  ,
\end{equation}
\\
is the usual dimensionless action calculated between the extremes
$\Phi_L (\varepsilon_0)$ and $\Phi_R (\varepsilon_0)$ given by
intersection of line $ \varepsilon =  \varepsilon_0$ with the
potential barrier. For a very little rotational perturbation ($h
\nu_\Omega \ll h \nu_{LC} $) the position of well bottoms remain
substantially unchanged, so that Eq. (\ref{19}) is still valid.
These quantities will be considered, for instance for the
following realistic values of an rf-SQUID: $L = 0.15 \;
\mathrm{nH}$, $I_c = 2.61 \; \mu\mathrm{ A}$, $C = 0.15 \;
\mathrm{pF}$,  $R = 1 \; \mu\mathrm{ m}$. With these values we
have, by numerical solution in symmetrical ``at rest" Hamiltonian:
\\
\begin{equation}\label{22}
\begin{array}{ccl}
  \Phi_1 & = & 0.1596 \; \Phi_0 \\\\
  \nu_{\mathrm{LC}} & \cong & 2.0 \times 10^{10} \; \mathrm{Hz} \\\\
  \nu_{\mathrm{tu}} & \cong & 1.59 \times 10^{5} \; \mathrm{Hz}
\end{array}
\end{equation}
\\
Because of the little perturbative effect to be considered by
rotation of rf-SQUID, these approximate values will be considered
in a ``rotating" case. The Hamiltonian frequency due to the
rotational (or Sagnac) shift for above values becomes
\\
\begin{equation}\label{23}
\nu_\Omega = 118610 \left( \frac{\Omega}{ \mathrm{rad} \;
\mathrm{s}^{-1}} \right).
\end{equation}
\\
 At a time period $T = 1/4 \nu_{\mathrm{tu}} $ the differential
probability for the unperturbed system [Eq. (\ref{8})] is zero
while the perturbed one [Eq. \ref{16}] gives, for any angular
velocity
 $\Omega> 0.5 \; \mathrm{rad} \; \mathrm{s}^{-1}$, values sensibly
greater than zero. For instance:
\\
\begin{equation}\label{24}
 \begin{array}{|c||c|c|} \hline
 P &   \Omega  \; (\mathrm{rad} \; \mathrm{s}^{-1})
   & h \nu_\Omega / \varepsilon_0  \times 10^{5}
   \\  \hline \hline
0.04 & 0.6 & .72 \\ \hline
  0.11 & 1 & 1.2 \\ \hline
   0.39 & 2 & 2.4 \\ \hline
    0.97 & 2 \pi & 7.6 \\ \hline
\end{array}
\end{equation}
\\
 This makes
realistic the experimental determination of the Mach principle at
a quantum level using a rf-SQUID mounted on a rotating platform.
Instead, it does not seem possible, at presently available
instrument sensitivity, to use the rf-SQUID  to determine the
Sagnac signature of the Earth rotation. Indeed the Eq. (\ref{16})
gives a not vanishing value of $P(t)$ only if the condition $
\nu_{\mathrm{tu}} \leq \nu_\Omega$ is satisfied. In order to avoid
dissipative effects on the wavefunction time evolution during the
measurement the tunnelling frequency must remain in the order of 1
MHz \cite{tes}. Of consequence the only possibility is that to
enhance $\nu_\Omega$ by increasing ring radius $R$, without any
changement of double well potential. For a SQUID mounted on a
rotating platform we have seen that $\Omega \cong 1$ rad/s and $R
= 1$ $\mu$m are realistic values. Examining Eq. (\ref{18}) it is
easily seen that the product ${\mathbf{\Omega}} \cdot
{\mathbf{\Sigma}}$ must remain the same for the case $\Omega =
\Omega_\oplus $, which implies $R \approx 110 $ $\mu$m, an
absolutely unrealistic value for today technology.
 We note that the stationary
rotating case may be reached adiabatically from a static (i.e.
non-rotating) case. Indeed  we may assume, as in the final table
(Eq. \ref{24}) a characteristic angular acceleration $\alpha \sim
\Omega / \mathrm{day}$, where $\Omega \approx 1 \;
\mathrm{rad/sec}$  which imply $\alpha / \Omega \ll \Omega \ll
\omega_{\mathrm{tu}} \ll \omega_{LC} $
 the quantum system
evolution may be described by a similar slow time variable
two-well potential from a static symmetric one toward the final
antisymmetric one.
\\
In conclusion we feel that it is exciting to suggest a first
realistic experimental opportunity to test a global cosmological
feature at microscopic quantum level.
%
%
%

%
%
\section*{Acknowledgment}
 We thank Prof. G. Diambrini and Prof. C. Cosmelli (responsible of the
 MQC experiment in Rome) and Dr. C. Bravi for continuous support and
 useful discussions; we are also deeply indebted to Dr. R. Conversano
 and Dr. A. Salis for reading the manuscript and useful comments.

\end{document}